\documentclass[a4paper,12pt,onecolumn]{article}
\usepackage{graphicx}
\usepackage[utf8]{inputenc}
\title{Formation of opinion of one's own conviction about a Small World topology}
\author{Victor H. Blanco\\Facultad de Ciencias Puras y Naturales - UMSA\\La Paz - Bolivia}

\begin{document}
\maketitle
\section{Abstract}
A model of opinion formation on a Small World type topology is studied to observe qualitatively its difference with one's own conviction versus the opinion of others. It is possible to observe that the relation between conviction to copy the agent with whom it interacts and this agent is an undecided agent who leaves decisions to chance becomes outstanding since the probability of undecided agents is greater than $0.1$ and is maintained until $0.9$ so that in the border regions, it is only a transition towards this predictive.
\section{Introduction}
In the social systems the Ising Model is essential its use, to be able to represent the interaction of the denominated agents (individuals in the society) and the interaction of some with others in many models described in \cite{caste} and other models with different dynamics like \cite{holyst,galam,banish,parsegov,dong,fan,ortiz,stern,li}, the model is quite used in variety of these used in the same topology not only opinion dynamics \cite{jiang,guzman,dall}, based on kinetic models of opinion, the population is based on graphs obtained from the program \textit{Pajek} versions where the option of construction of the adjacent matrix with characteristics of network type Small World can be found, as the networks of this type are variable, a network of size $600$ is added and from this two others of smaller size $300$ and $100$ are extracted, hoping that characteristics of the base model can be found for this work.
\section{Modelo}
Continuing with the model used by \cite{nuno}, we would now represent the number of undecided agents against the consensus of the population regardless of the number of the population, so the model used would be similar:
\begin{itemize}
\item We randomly choose a pair of agents {i,j} for the interaction.
\item If agent $i$ represents an undecided agent, where he considers his own opinion as valuable or insignificant to the opinion $j$ denoted random.
\item The iteration relation is:
\begin{displaymath}
o_{1}=Ao_{i}+\mu_{ij}o_{j}
\end{displaymath}
Where the parameter $A$ represents the agent's own conviction to his previous decision, the matrix $mu_{ij}$ represents the response quality of agent $j$ and $o_{j}$ is the opinion of agent $j$ is momentum.\\
Where this process is repeated $N$ times depending on the size of the selected population.\\
\end{itemize}
It can be noted that the matrix $\mu_{ij}$ can represent positive or negative with probability $p$ or $1-p$, in other words the randomness of the interacting agent is added to the indeterminacy of the first agent, which adds more contradiction to the decision to be made about the interacting agents.\\
\section{Results}
\begin{figure}[!ht]
\begin{center}
 \includegraphics[scale=0.55]{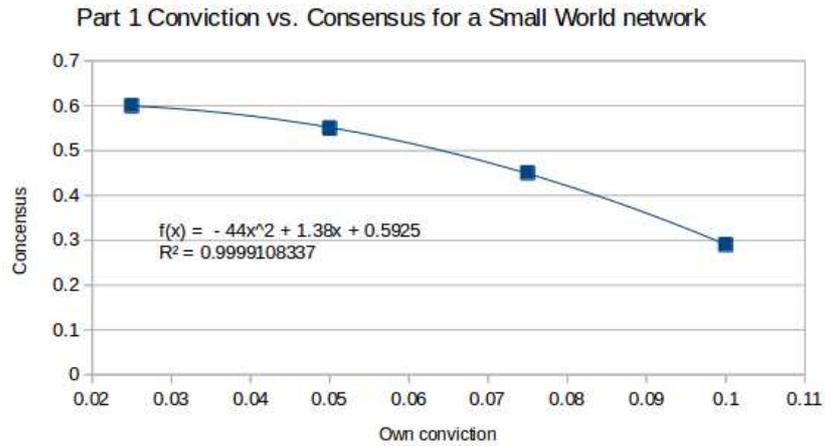}
 \caption{Trend between $0$ to $0.1$ of agents with conviction.}
 \label{1raparte}
\end{center}
\end{figure}

\begin{figure}[!ht]
\begin{center}
 \includegraphics[scale=0.55]{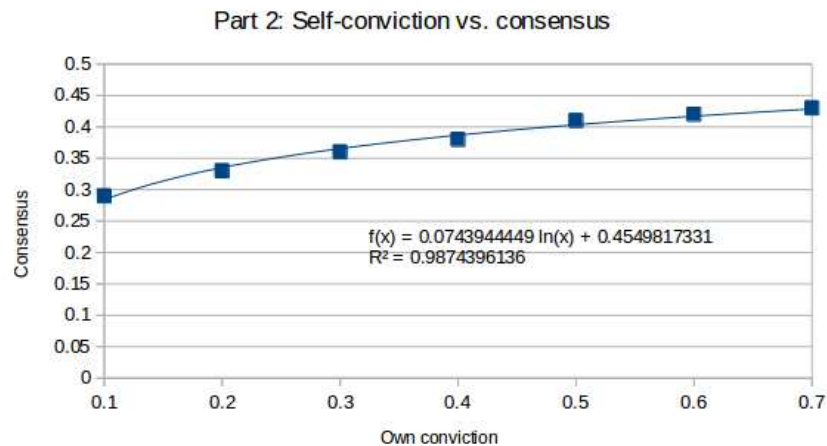}
 \caption{Trend between $0.1$ and $0.7$ of agents with conviction.}
 \label{2daparte}
\end{center}
\end{figure}

For the "quenched" case, the figure is obtained (\ref{2dapart}):
\begin{figure}[!ht]
\begin{center}
 \includegraphics[scale=0.9]{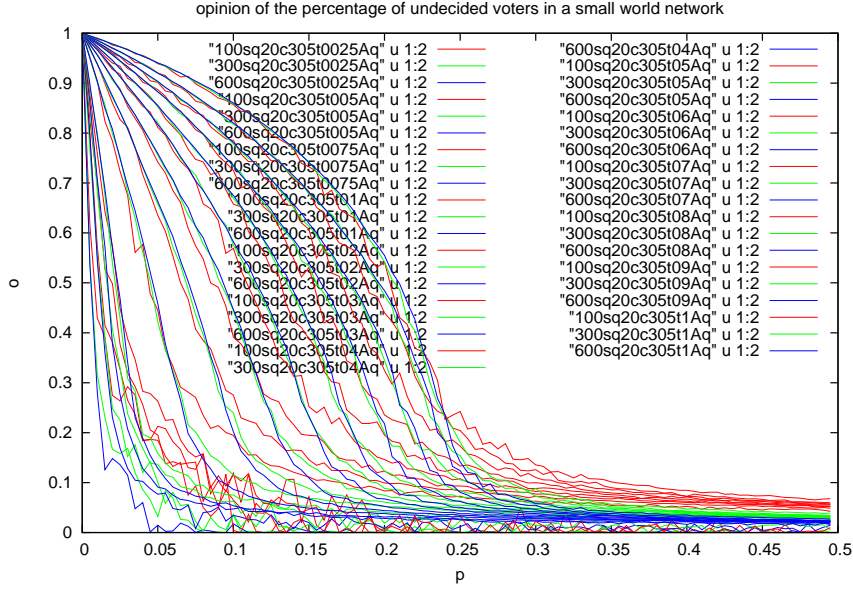}
 \caption{Trend between $0$ and $1$ of agents with conviction with different population.}
 \label{2dapart}
\end{center}
\end{figure}
It can be observed that the smaller population of 100 individuals (red lines in \ref{2dapart}), in all cases is the one that does not reach decision convergence and for more populated sets the decision is more defined, which leads us to divide these results into two parts, one of $A$ conviction of the agents of
\begin{enumerate}
\item Part 1: for $0.1$ a $0.1$
\item Part 2: More to $0.1$
\end{enumerate}
For the quenched case, but knowing that there are more update modes shown in \cite{nuno} where they are shown as equivalent, we obtain the figures \ref{1raparte} and \ref{2daparte}:
Case between 0 and 0.1, we obtain the relation of $C_p$:own conviction and $c$:consensus:
\begin{equation}
C_p=-44c^2+1.38c+0.592 \in[0;0,1]
\end{equation}
shown in figure(\ref{1raparte}) with correlation coefficient $r_1=0.998$.\\
For the second part, where the relationship that emerges is one approximating a linear relationship of the type:
\begin{equation}
C_p=0.267c+0.272 \in[0,1;0,7]
\end{equation}
With a correlation coefficient of $R=0.992$.\\
The theoretical resolution is the one that approaches the section of part 1, since being random the conviction would correspond to a function as the identity starting from $0.25$ of the population and affecting a quarter of this one obtaining a probability of conversion of $0.5$ for the total of the population, it is evident that an abrupt transition cannot exist and this is found at the beginning of the graph as at the end and is valid for part 2 only, where most of the behavior emerges, of conviction copying undecided.\\
Part 2, As the trend is close to $0.1$ from the $0.7$ conviction it is equivalent to say that from this value the choice will be random with no preferences.\\
\section{Conclusions}
The consensus obtained from agents in a small-world network shows two distinct sections, one with a parabolic inverse trend between [0;0.1] (part 1) downward trend and the other with a linear trend between [0.1;0.7] (part 2), delimiting the consensus by a quadratic line with upward trend, all results with parameter the consensus.

\end{document}